\documentclass[11pt]{article}

\usepackage[english]{babel}

\usepackage{amssymb,epsf,epsfig}
\usepackage{amsmath}

\usepackage{graphicx}

\usepackage{slashed}
\usepackage[active]{srcltx}

\def\II{\hbox{{1}\kern-.25em\hbox{l}}}

\newcommand{\z}{\underline{z}}

\newcommand \VEV [1] {\left\langle{#1}\right\rangle}
\newcommand \vev [1] {\langle{#1}\rangle}

\newcommand{\ms}{\mskip 1.5mu}

\newcommand{\im}{\operatorname{Im}}

\newcommand{\beq}[1]{
\begin{equation}\label{#1}}

\newcommand{\eeq}{\end{equation}}

\newcommand{\bea}[1]{
\begin{eqnarray}\label{#1}}

\newcommand{\eea}{\end{eqnarray}}

\begin{document}

\begin{center}
\textbf{\LARGE
Integrability of the evolution equations for heavy-light baryon distribution amplitudes}

\vspace*{1.6cm}

{\large V.M.~Braun$\ms{}^1$, S.E.~Derkachov$\ms{}^{2,3}$  and  A.N.~Manashov$\ms{}^{1,4}$}

\vspace*{0.4cm}

\parbox[t]{0.93\textwidth}{\small\it
$^1$ Institut f\"ur Theoretische Physik, Universit\"at Regensburg,
D-93040 Regensburg, Germany \\
$^2$ St.~Petersburg Department of Steklov Mathematical Institute, 191023 St.-Petersburg, Russia\\
$^3$ St.~Petersburg State Polytechnic University, 195251 St.~Petersburg,  Russia\\
$^4$ Department of Theoretical Physics,  St.-Petersburg University, 199034 St.-Petersburg, Russia}

\vspace*{0.8cm}

\textbf{Abstract}\\[10pt]
\parbox[t]{0.85\textwidth}{\small
We consider evolution equations describing the scale dependence of the wave function
of a baryon containing an infinitely heavy quark and a pair of light quarks at small transverse
separations, which is the QCD analogue of the helium atom. 
The evolution equations depend on the relative helicity of the light quarks.
For the aligned helicities, we find that the equation is completely integrable, that is 
it has a nontrivial integral of motion, and 
obtain exact analytic expressions for the eigenfunctions and the anomalous dimensions. 
The evolution equation for anti-aligned helicities contains an extra term that breaks integrability and 
creates a ``bound state'' with the anomalous dimension separated from the rest of the spectrum by a finite gap.
The corresponding eigenfunction is found using numerical methods.
It describes the momentum fraction distribution of the light quarks in, e.g., $\Lambda_b$-baryon
at large scales.
}

\date{\today}

\end{center}

\vspace*{1cm}





%
%

{\large \bf 1.}~~Precision tests of the flavor sector of the Standard Model may reveal new physics and remain high
on the agenda. Main attention has been so far focused on B-mesons but interest is developing to the heavy baryon
decays as well. Such baryons are produced copiously at the LHC and, as more data are collected, studies of rare
$b$-baryon decays involving flavor-changing neutral current transitions have to become quantitative in order  to
make an impact on the field. In particular the $\Lambda_b\to \Lambda \mu^+\mu^-$ decays are receiving a lot
of attention, see e.g. Ref.~\cite{Feldmann:2011xf} and references therein.

Theoretical description of the $b$-hadron decays is based on
factorization theorems
that make use of the large mass of the $b$-quark in order to separate calculable effects
of short distances from the nonperturbative large distance physics.
The corresponding formalism is similar but much less developed for baryons as compared to mesons.
A recent discussion using SCET formalism can be found in Ref.~\cite{Wang:2011uv}.
For the exclusive decays involving large energy release in the final state, the
relevant nonperturbative quantities are baryon wave functions at  small transverse separations,
dubbed light-cone distribution amplitudes~(DA). Their study was started in
Refs.~\cite{Ball:2008fw,Ali:2012pn,Bell:2013tfa} where the complete
classification and renormalization group equations (RGE) that govern the scale-dependence are presented.

In this work we point out that these equations have a hidden symmetry and completely integrable in the case that the
light quarks have the same helicity. In other words, we identify a nontrivial quantum number that distinguishes
heavy baryon states with different scale dependence and obtain exact analytic solution of the evolution equations.
This phenomenon is similar to integrability of RGEs  for the light baryons
\cite{Braun:1998id,Braun:1999te} and, in a more general context, to integrability in high-energy QCD
~\cite{Lipatov:1993yb,Lipatov:1994xy,Faddeev:1994zg} and  in the $N=4$ supersymmetric Yang-Mills
theory~\cite{Minahan:2002ve,Beisert:2003tq,Beisert:2003yb} that attracted huge attention as a tool to check the
AdS/CFT correspondence. The integrable model that we encounter in the present context is different; it has been
discussed recently in~\cite{Derkachov:2014gya}.

The similar equation for the case that the two light quarks have opposite helicity
contains an extra term that breaks integrability and creates a ``bound state''
with the anomalous dimension separated from the rest of the spectrum by a finite gap.
The corresponding eigenfunction is found numerically.
It describes the momentum fraction distribution of the light quarks in, e.g. $\Lambda_b$,
at large scales and can be called ``asymptotic DA'' in analogy to the accepted terminology for hadrons built of light quarks.

%
\vspace*{0.5cm}
%

{\large \bf 2.}~~Consider at first the leading-twist DA of a baryon containing an infinitely heavy
quark and a transversely polarized ``diquark'': a pair of light quarks with aligned helicities.
It can be defined as~\cite{Ali:2012pn}
\begin{eqnarray}
\langle 0| [q^T_1(z_1 n)C \slashed{n}\gamma^\mu_\perp q_2(z_2 n)]h_v(0)|B^{j=1}(v)\rangle
&=&
 \frac{1}{\sqrt{3}} \epsilon^\mu_\perp u(v)\,  f^{(2)}_B(\mu) \Psi_\perp(z_1,z_2;\mu)\,.
\label{def:DA}
\end{eqnarray}
Here $q_{1,2} = u,d,s$ are light quarks separated by a lightlike distance,
$h_v(0)$ is the effective heavy quark field with four-velocity $v$,
$C$ is the charge conjugation matrix, $u(v)$ is the Dirac spinor $\slashed{v}u(v) = u(v)$,
and $\epsilon^\mu$ is the diquark polarization vector, $v^\mu \epsilon_\mu=0$.
The transverse projections are defined with respect to the two auxiliary light-like vectors
$n$ and $\bar n$ which we choose such that $v_\mu = (n_\mu+\bar n_\mu)/2$, $v\cdot n =1$,
$n\cdot\bar n=2$:
\begin{align}
   \epsilon^\mu_\perp = g^{\mu\nu}_\perp \epsilon_\nu\,, &&
   g^{\mu\nu}_\perp = g^{\mu\nu} - (n^\mu \bar n^\nu +n^\nu \bar n^\mu)/(n\cdot\bar n)\,,
\end{align}
and similar for $\gamma^\mu_\perp$. The Wilson lines connecting the quark fields are
not shown for brevity. The heavy quark field $h_v$ can itself be related to the
Wilson line as~\cite{Korchemsky:1991zp}
\begin{equation}
       \langle 0| h_v(0) |h,v\rangle = {\rm Pexp}\left[ig\int_{-\infty}^0\!d\alpha\,v_\mu A^\mu(\alpha v)\right]\,,
\end{equation}
so that the operator in Eq.~(\ref{def:DA}) can be viewed as a pair of light quarks (a diquark), attached to the
Wilson line with a cusp containing one  lightlike and one timelike segment.
Finally, the coupling $f^{(2)}_B$ is defined as the
matrix element of the corresponding local $q_1 q_2 h_v$ operator; it is inserted for
normalization~\cite{Ali:2012pn}. The parameter $\mu$ is the renormalization (factorization) scale. We tacitly imply
using $\overline{MS}$ scheme.

The product $\epsilon^\mu_\perp u(v)$ on the right-hand-side (r.h.s.) of Eq.~(\ref{def:DA})
can be expanded in irreducible representations corresponding to physical baryon states
with $J^P = 1/2^+$ and $J^P = 3/2^+$ using suitable projection operators, see~\cite{Ali:2012pn}.
These (ground) states form the $SU(3)_F$ multiplets (sextets), $\Sigma_b,\Xi_b,\Omega_b$ and
$\Sigma^\ast_b,\Xi^\ast_b,\Omega^\ast_b$, respectively, which are degenerate in the heavy $b$-quark limit.
The double-strange $\Omega_b$ baryon is of special interest for flavor physics as it only
decays through weak interaction. The DA $\Psi_\perp(z_1,z_2;\mu)$
is written usually in terms of its Fourier transform
\begin{align}
 \Psi_\perp(\z;\mu) &= \int_0^\infty\!\!\! d\omega_1 \! \int_0^\infty\!\!\! d\omega_2 \,
  e^{-i(\omega_1 z_1 + \omega_2 z_2)}\Psi_\perp(\omega_1,\omega_2;\mu)\,
= \int_0^\infty\!\!\!\omega d\omega\!\int_0^1\!\!\! du e^{-i\omega(uz_2+\bar u z_1)}\, \widetilde \Psi_\perp(\omega,u;\mu)\,,
\label{DA1}
\end{align}
where $\z =\{z_1,z_2\}$
and in the second line we redefine $\omega_1 = u \omega$, $\omega_2 =  \bar u \omega$ with $\bar u =1-u$.
The variables $\omega_1$ and $\omega_2$ correspond to the energies (up to a factor two)
of light quarks in the baryon  rest frame. The DA (\ref{DA1}) 
is the most important 
nonperturbative input in QCD calculations of exclusive heavy baryon decays to the leading power 
accuracy in the heavy quark mass.

The scale dependence of $\widetilde \Psi_\perp(\omega,u,\mu)$ or, equivalently, $\Psi_\perp(\underline{z};\mu)$,
is governed by the renormalization group equation~\cite{Ball:2008fw,Ali:2012pn}
\begin{align}\label{RGE}
 \left(\mu\frac{\partial}{\partial\mu}+\beta(\alpha_s) \frac{\partial}{\partial \alpha_s}+
  \frac{2\alpha_s}{3\pi}\mathbb{H}\right) f^{(2)}_B(\mu)\,\Psi_\perp(\z;\mu)=0\,.
\end{align}
The evolution kernel $\mathbb{H}$ is an integral operator which can be decomposed as
\begin{align}\label{HH}
\mathbb{H} = \mathcal{H}_{12}+\mathcal{H}^h_1+\mathcal{H}^h_2-4\,.
\end{align}
The kernels $\mathcal{H}_k^h$ are due to heavy-light quark interactions,
\begin{align}
\mathcal{H}^h_1f(\z)&=\int_0^1\frac{d\alpha}{\alpha}\Big[f(\z)-\bar\alpha f(\bar\alpha z_1,z_2)\Big]
+ \ln (iz_1 \mu)\,f(\z)\,,
\notag\\
\mathcal{H}^h_2f(\z)&=\int_0^1\frac{d\alpha}{\alpha}\Big[f(\z)-\bar\alpha f(z_1,\bar\alpha z_2)\Big]
 + \ln (iz_2 \mu)\,f(\z)\,.
\label{heavy}
\end{align}
They are identical to the Lange-Neubert kernels~\cite{Lange:2003ff,Braun:2003wx,Knodlseder:2011gc}.
The remaining contribution
\begin{align}
{\mathcal{H}}_{12}f(\z)&=
\int_0^1\frac{d\alpha}{\alpha}\Big[2f(\z)-\bar\alpha f(z_{12}^\alpha,z_2)
- \bar\alpha f(z_1,z_{21}^\alpha))\Big]
\label{light}
\end{align}
takes into account the interaction between the light quarks; it is similar to the standard
Efremov-Radyushkin-Brodsky-Lepage evolution kernel for the pion DA.
Here and below we use the notation
\begin{align}
 z_{12}^\alpha = \bar\alpha z_1 + \alpha z_2\,, && \bar \alpha = 1-\alpha\,.
\label{lignt}
\end{align}
The evolution kernels (\ref{heavy}), (\ref{light}) can be written in terms of the generators of
the collinear subgroup of conformal transformations
\begin{align}
  S_+ = z^2 \partial_z + 2j z \,, && S_0 = z\partial_z + j\,, && S_- = -\partial_z\,,
\label{generators}
\end{align}
where $j=1$ is the conformal spin of the light quark.
The generators satisfy the standard $SL(2)$ commutation relations
\begin{equation}
  [S_+,S_-] = 2 S_0\,, \qquad [S_0,S_\pm] = \pm S_\pm\,.
\label{SL2}
\end{equation}
One can show that~\cite{Braun:2014owa}
\begin{align*}
 \mathcal{H}^h_1 = \ln\big(i\mu S_+^{(1)}\big)-\psi(1)\,,
&&
 \mathcal{H}^h_2 = \ln\big(i\mu S_+^{(2)}\big)-\psi(1)\,,
\end{align*}
where $S_+^{(1)}$, $S_+^{(2)}$ act on the first, $z_1$, and the second, $z_2$, light-cone coordinate, respectively. The
last kernel (\ref{light}) is written in terms of the two-particle Casimir operator $S_{12}^2$~\cite{Bukhvostov:1985rn}
\begin{align}
{\mathcal{H}}_{12} = 2\big[\psi(J_{12})-\psi(1)\big]\,,
\end{align}
where
$
   S_{12}^2 = S_+ S_- + S_0(S_0-1)= J_{12}(J_{12}-1)
$, $S_+ = S_+^{(1)}+ S_+^{(2)}$ etc., 
and $\psi(x)$ is the Euler's digamma function.
Thus, the complete evolution kernel takes a very compact form
\begin{align}
 \mathbb{H} &= \ln\big(i\mu S_+^{(1)}\big) + \ln\big(i\mu S_+^{(2)}\big)
   + 2\psi(J_{12})-4 \psi(2)\,.
\label{HHH}
\end{align}
The evolution equation for the DA in momentum space, $\Psi_\perp(w_1,w_2;\mu)$,
is given by the same expression with the $SL(2)$ generators in the momentum space representation~\cite{Braun:2003rp}.
Eigenfunctions of $\mathbb{H}$ correspond to the states that have autonomous
scale dependence and the corresponding eigenvalues define anomalous
dimensions.

%
\vspace*{0.5cm}
%

{\large \bf 3.}~~Our main result is that this evolution equation can be solved explicitly.
To this end  we consider the following operators
\begin{align}
 \mathbb{Q}_1 = i\big(S_+^{(1)}+S_+^{(2)}\big),
&&
 \mathbb{Q}_2 = S_0^{(1)}S_+^{(2)}-S_0^{(2)}S_+^{(1)}.
\end{align}
It is possible to show that $\mathbb{Q}_1$ and $\mathbb{Q}_2 $ commute with each other
and with the evolution kernel~$\mathbb{H}$:
\begin{align}\label{QHcomm}
{}[\mathbb{Q}_1,\mathbb{Q}_2] = [\mathbb{Q}_1,\mathbb{H}] = [\mathbb{Q}_2,\mathbb{H}] = 0\,.
\end{align}
The first two relations are trivial, the last one can be verified
using the explicit expressions for $\mathbb{Q}_2$ and $\mathbb{H}$.

If $\mathbb{H}$ is interpreted as a Hamiltonian of a certain
quantum-mechanical model, the operators $\mathbb{Q}_1$ and
$\mathbb{Q}_2$ correspond to the conserved charges.
In the formalism of the quantum inverse scattering method (QISM) the charges $\mathbb{Q}_1, \mathbb{Q}_2$
appear in the expansion of the element $C(u)$ of the monodromy matrix,
$$C(u)=u\,\mathbb{Q}_1+\mathbb{Q}_2.$$ The commutation relation $[C(u),\mathbb{H}]=0$, and its generalization
to more degrees of freedom then follow directly from the QISM~\cite{Faddeev:1996iy,Kulish:1981bi,Sklyanin:1991ss}.
Note that in classical applications of integrable models one encounters Hamiltonians that commute with the sum of
diagonal elements, $A(u)+D(u)$, of the monodromy matrix. In our case the Hamiltonian commutes with $C(u)$, which
corresponds to a new, nonstandard integrable model.

The conserved charges $\mathbb{Q}_1$, $\mathbb{Q}_2$ are self-adjoint operators with respect to
the $SL(2,R)$ invariant  scalar product
\begin{align}\label{sc}
\VEV{\Phi|\Psi}=\frac{1}{\pi^2}\int_{\mathbb{C}_-} d^2z_1\int_{\mathbb{C}_-}d^2z_2\,\big(\Phi(\underline{z})\big)^*\,\Psi(\underline{z})\,,
\end{align}
where the integration goes over the lower complex half-plane, $\im z_i<0$. The eigenfunctions  of $C(u)$ provide the
basis of the so-called Sklyanin's representation of Separated Variables and are known in explicit form~\cite{Derkachov:2002tf}.
They are labeled, for the present case, by two real numbers: $s>0$ and $x\in \mathbb{R}$ such that
\begin{align}\label{CPsi}
C(u)\phi_{s,x}(z_1,z_2)=s(u-x)\phi_{s,x}(z_1,z_2)\,,
\end{align}
with
\begin{align}\label{Psidef}
\phi_{s,x}(\underline{z})&=\frac{s}{z_1^2 z_2^2}\int_0^1d\alpha\,\left(\frac{\alpha}{\bar\alpha}\right)^{ix}
\exp\big[is(\bar\alpha/z_1+\alpha/z_2)\big]
=
s\rho(x)\frac{e^{is/z_1}}{z^2_1z_2^2}\,{}_1F_1\Big(1\!+\!ix,2,is \big(z^{-1}_2\!-\!z^{-1}_1\big)\Big),
\end{align}
where
\begin{align}
  \rho(x)=\pi x/\sinh(\pi x).
\end{align}
The eigenfunctions $\phi_{s,x}(\underline{z})$ form a complete system in the Hilbert space defined
by the scalar product~(\ref{sc})
\begin{align}
\VEV{\phi_{s',x'}|\phi_{s,x}}= \frac{2\pi}{s}\, \delta(s-s')\delta(x-x')\,.
\end{align}
Since the conserved charges $\mathbb{Q}_1$ and $\mathbb{Q}_2$ commute with the Hamiltonian $\mathbb{H}$,
they share the same  set of eigenfunctions,
\begin{align}\label{HPsi}
\mathbb{H}\,\phi_{s,x}(\underline{z})={\gamma}(s,x)\,\phi_{s,x}(\underline{z})\,.
\end{align}
The simplest way to calculate the eigenvalues is to compare the large-$z$  asymptotics of the expressions on the both sides
of this equation. In this way one obtains the anomalous dimensions
\begin{align}
{\gamma}(s,x;\mu)=2\ln(\mu s/s_0)+\mathcal{E}(x)\,, &&  \mathcal{E}(x)= \psi(1+i x)+\psi(1-i x)+2\gamma_E
\label{gamma}
\end{align}
where $s_0=e^{2-\gamma_E}$.
Going back to the RGE equation (\ref{RGE}), we expand the DA $\Psi_\perp(\underline{z},\mu)$ over the
eigenfunctions of $\mathbb{H}$
\begin{align}
\Psi_\perp(\underline{z},\mu)
=\int_0^\infty ds\,s\int_{-\infty}^{\infty}\frac{dx}{2\pi}\, \eta_\perp(s,x;\mu)\,\phi_{s,x}(\underline{z})\,.
\end{align}
The expansion coefficients
$
\eta_\perp(s,x;\mu)=\VEV{\phi_{s,x}|\Psi_\perp}
$
evolve autonomously,
\begin{align}
f^{(2)}_B(\mu)\eta_\perp(s,x;\mu)&=f^{(2)}_B(\mu_0)\eta_\perp(s,x;\mu_0)\,\left(\frac{\mu}{\mu_0}\right)^{-\frac{8}{3\beta_0}}
\left(\frac{\mu_0 s}{s_0}\right)^{\frac{8}{3\beta_0}\ln L}\,L^{\frac{4}{3\beta_0}
\left[\mathcal{E}(x)-\frac{4\pi}{\beta_0\,\alpha_s(\mu_0)}\right]},
\label{scaledep}
\end{align}
where
$L = {\alpha_s(\mu)}/{\alpha_s(\mu_0)}$,  $\beta_0=\frac{11}3N_c-\frac23 n_f$.
For large scales, the coefficients  $\eta_\perp(s,x;\mu)$ slowly drift towards smaller values of both
parameters: $s\to 0$, thanks to the factor $s^{8\ln L /3\beta_0}$, and $|x|\to 0$, taking into account that
$\psi(1+ix)\sim \ln |x|$ for $x\to\pm\infty$.

Going over to the DA in momentum space, $\widetilde\Psi_\perp(\omega,u;\mu)$, we define the corresponding eigenfunctions
as
\begin{align}
\widetilde\phi_{s,x}(\omega,u) = \vev{e^{-i \omega(\bar uz_1+ u z_2)}|\phi_{s,x}(z_1,z_2)}\,.
\end{align}
Using that $\langle e^{-ikz}|z^{-2}e^{is/z}\rangle = - (1/\sqrt{ks}) J_1(2\sqrt{ks})$~\cite{Braun:2014owa} one obtains
\begin{eqnarray}
\widetilde\phi_{s,x}(\omega,u)&=&\frac1{\omega\sqrt{u\bar u}}\int_0^1\frac{d\alpha}{\sqrt{\alpha\bar\alpha}}\alpha^{ix}\bar\alpha^{-ix}\,
J_1\left(2\sqrt{ws\bar\alpha\bar u}\right)J_1\left(2\sqrt{ws\alpha u}\right)\,.
\label{eigenfunctions}
\end{eqnarray}
The eigenfunctions $\widetilde\phi_{s,x}(\omega,u)$ are orthogonal and form a complete set:
\begin{align}
&\frac{s}{2\pi} \int_0^\infty \omega^3 d\omega \int_0^1 du\,u\bar u\,
\widetilde\phi_{s,x}(\omega,u)\, \widetilde\phi^\ast_{s',x'}(\omega,u)=\delta(s-s')\delta(x-x')\,,
\\[2mm]
& \omega^3u\bar u \int_0^\infty s ds \int_{-\infty}^\infty \frac{dx}{2\pi} \,
\widetilde\phi_{s,x}(\omega,u)\, \widetilde\phi^\ast_{s,x}(\omega',u')
={\delta(\omega-\omega')\delta(u-u')}\,.
\end{align}
Making use of the Bateman's expansion for the product of two Bessel functions we obtain a series representation
\begin{eqnarray}
\widetilde\phi_{s,x}(\omega,u)&=&\frac1{\omega}\sum_{n=0}^\infty i^n \varkappa_n^{-1}\, C_n^{3/2}(1-2u)\, H_n(x)\,\frac{1}{\sqrt{s\omega}}J_{2n+3}(2\sqrt{s\omega})\,.
\end{eqnarray}
Here $C^{3/2}_n(x)$ are the Gegenbauer polynomials, $J_{2n+3}(x)$ are Bessel functions, and
\begin{align}
 \varkappa_n &= \frac{(n+1)(n+2)}{4(2n+3)}\,.
\end{align}
The functions $H_n(x)$ are given by the continuous Hahn polynomials up to the
prefactor $\rho(x)$:
\begin{align}
H_n(x)=i^n \int_0^1 du \left(\frac{u}{\bar u}\right)^{ix} \, C_n^{3/2}(1-2u)=
\frac{(n+1)(n+2)}{2} i^n\rho(x)\,{}_3F_2\Big(\genfrac{}{}{0pt}{}{-n,n+3,1+ix}{2,2}\Big|1\Big)\,,
\end{align}
e.g.
\begin{align}
\rho^{-1}(x)H_0(x)= 1\,, && \rho^{-1}(x)H_1(x)= 3 x\,, && \rho^{-1}(x)H_2(x)= 5x^2-1\,,
\end{align}
etc.
Hahn polynomials are real functions, have the symmetry $H_n(x)=(-1)^n H_n(-x)$,
and form a complete orthogonal system. In our normalization
\begin{align}
\int_{-\infty}^\infty \frac{dx}{2\pi} H_n(x)\, H_m(x)= \varkappa_n\, \delta_{mn}\,.
\end{align}
Collecting everything we obtain the final result:
\begin{align}
 \widetilde \Psi_\perp(\omega,u;\mu)&= \omega^2 u\bar u \int_{-\infty}^\infty\! \frac{dx}{2\pi}\int_0^\infty\!\! s ds\,
 \widetilde\phi_{s,x}(\omega,u) \eta_\perp(s,x;\mu)\,,
\end{align}
where the scale dependence of $\eta_\perp(s,x;\mu)$ is given in Eq.~(\ref{scaledep}).
The expansion coefficients of $\eta_\perp(s,x;\mu)$ in Hahn polynomials are related to the expansion 
coefficients of $\widetilde \Psi_\perp(\omega,u;\mu)$ in Gegenbauer polynomials,
\begin{align}
\eta_\perp(s,x;\mu) = \sum_n i^n \,\eta_{n}^\perp(s;\mu) H_n(x) \qquad\mapsto\qquad
&& \widetilde \Psi_\perp(\omega,u;\mu) =  u\bar u \sum_n \psi^\perp_n(\omega;\mu)\, C_n^{3/2}(2u-1)\,,
\label{b2}
\end{align}
by the Bessel transform (cf. Eq.~(30) in Ref.~\cite{Braun:2014owa})
\begin{align}
\psi^\perp_n(\omega;\mu)=\int_0^\infty ds \sqrt{s\omega}\, J_{2n+3}\left(2\sqrt{s\omega}\right)\, \eta^\perp_n(s;\mu)\,.
\label{b1}
\end{align}
Making use of the asymptotic expansion for the Bessel function one finds that the small-$s$ behavior 
$\eta^\perp_n(s)\sim s^{p_n}$ translates into the large-$\omega$ asymptotics of the function 
$\psi^\perp_n(\omega)\sim \omega^{-1-p_n}$ unless there is some cancellation, see below.

The expansion coefficients at the reference (low) scale can
be calculated from a given model of the DA as
\begin{align}
 \eta_\perp(s,x;\mu_0) = \int_0^\infty \omega d\omega \int_0^1 du\,
 \widetilde\phi^\ast_{s,x}(\omega,u)\,\widetilde \Psi_\perp(\omega,u;\mu_0).
\end{align}
In the existing studies it is usually assumed that $\widetilde \Psi_\perp(\omega,u;\mu_0)$ is decreasing 
exponentially at large energies $\omega$. For a rather general model of this type
\begin{align}
 \widetilde \Psi_\perp(\omega,u;\mu_0) &= \omega^2 u\bar u \sum_n c_n
  \left(\frac{\omega}{\epsilon_n}\right)^{\kappa_n}\frac{e^{-\omega/\epsilon_n}}{\epsilon_n^4} C_n^{3/2}(2u-1)
\label{mod1}
\end{align}
one obtains
\begin{align}
\eta_\perp(s,x;\mu_0) &= {s}\sum_n {i^n c_n(s\epsilon_n)^{n}} H_n(x)
 \frac{\Gamma(n+4+\kappa_n)}{ \Gamma(2n+4)}
{}_1F_1\genfrac{(}{|}{0pt}{0}{n+4+\kappa_n}{2n+4}-s\epsilon_n\biggr).
\end{align}
In particular, for the simplest phenomenologically acceptable model~\cite{Ball:2008fw,Ali:2012pn,Bell:2013tfa}
\begin{align}\label{etasx}
  \widetilde \Psi_\perp(\omega,u;\mu_0) ~=~ \omega^2 u\bar u \, \frac{e^{-\omega/\epsilon_0}}{\epsilon_0^4}
\qquad\mapsto\qquad
   \eta_\perp(s,x;\mu_0) ~=~  \rho(x) \,s\,e^{-s\epsilon_0}\,.
\end{align}
Exponential decrease $\sim e^{-\omega/\epsilon_n}$ of each Gegenbauer harmonics in (\ref{mod1})
amounts, from the view point of the relation in Eq.~(\ref{b1}), to the fine tuning such that 
all leading power terms in the asymptotics $\omega \to \infty$ drop out. 
This fine tuning is, however, destroyed by the evolution so that a power-like asymptotics
is always generated.    

To see this, consider the simplest model in (\ref{etasx}) corresponding to the term $n=0$ in (\ref{mod1}) 
as an example. As the result of the evolution (\ref{scaledep}) all harmonics with $n>0$ become excited 
\begin{align}
\eta^\perp_n(s,\mu)=c_n(\mu)\, s (\mu_0s)^{-\delta}\, e^{-s\epsilon_0}
\end{align}
where $\delta=-{8}/{3\beta_0}\ln L$ and
\begin{align}
c_n(\mu)\sim \int dx\, H_0(x)\, L^{4/3\beta_0 \mathcal{E}(x)}\, H_n(x)\,.
\end{align}
For the corresponding coefficients in the Gegenbauer expansion (\ref{b2}) one obtains using (\ref{b1})
\begin{align}
\psi^\perp_n(\omega,\mu)=
c_n(\mu)\,\epsilon_0^{-2} \left(\frac{\epsilon_0}{\mu_0}\right)^\delta \left(\frac{\omega}{\epsilon_0}\right)^{n+2}
\frac{\Gamma(n+4-\delta)}{\Gamma(2n+4)}{}_1F_1\genfrac{(}{|}{0pt}{0}{n+4-\delta}{2n+4}-\frac{\omega}{\epsilon_0}\biggr).
\end{align}
The confluent hypergeometric function ${}_1F_1(a,b|\omega)$ decreases as a power of $\omega$ 
at $\omega\to \infty$, cf. Eq.~(\ref{1F1}) below, unless $a-b$ is a nonnegative integer,
in which case the asymptotic behavior is exponential. Thus, unless $\delta=0$ and $n=0$, we obtain
\begin{align}
\psi^\perp_n(\omega,\mu)\sim (\omega/\epsilon_0)^{-2+\delta}\,.
\end{align}
Note that the asymptotic behavior is the same for any $n$.

%
\vspace*{0.5cm}
%

{\large \bf 4.}~~Next, we consider heavy baryons with the light quarks having opposite helicity.
The scale dependence of the leading twist DAs does not depend on the spin of the light quark pair
and is the same for the $j^P=0^+$ $SU(3)_F$ triplet and all longitudinal DAs of heavy baryons in the
$j^P=1^+$ sextets, see~\cite{Ali:2012pn}. For definiteness, consider the $\Lambda_b$-baryon DA~\cite{Ball:2008fw,Ali:2012pn} defined as
\begin{eqnarray}
\langle 0| [u^T(z_1 n)C \gamma_5 \slashed{n} d(z_2 n)]h_v(0)|\Lambda(v)\rangle
&=&
f^{(1)}_\Lambda(\mu) \Psi_\Lambda(z_1,z_2;\mu)\,u_\Lambda(v)\,.
\label{def:LambdaDA}
\end{eqnarray}
The evolution equation for $\Psi_\Lambda(z_1,z_2;\mu)$ contains an additional term
corresponding to the gluon exchange between the
light quarks (in Feynman gauge)
\begin{align}
 \mathcal{H}_{12} \mapsto \mathcal{H}_{12} - \delta \mathcal{H}_{12}\,,  &&
\delta\mathcal{H}_{12}f(\z) =  \int_0^1d\alpha \int_0^{\bar\alpha}d\beta f(z_{12}^\alpha,z_{21}^\beta)
\end{align}
that corresponds to $\mathbb{H}\mapsto \mathbb{H}-1/J_{12}(J_{12}-1)$ in the $SL(2)$-invariant representation
of the evolution kernel in Eq.~(\ref{HHH}).
Expanding $\Psi_\Lambda(z_1,z_2;\mu)$ in terms of the eigenfunctions (\ref{Psidef}) of the integrable Hamiltonian (\ref{HHH})
\begin{align}
\Psi_\Lambda(\underline{z},\mu)
=\int_0^\infty ds\,s\int_{-\infty}^{\infty}\frac{dx}{2\pi}\, \eta_\Lambda(s,x;\mu)\,\phi_{s,x}(\underline{z})\,
\label{expandLambda}
\end{align}
one obtains the RGE equation for the expansion coefficients $\eta_\Lambda(s,x,\mu)$
\begin{multline}\label{RGE-V}
\left(\mu\frac{\partial}{\partial\mu}+\beta(\alpha_s) \frac{\partial}{\partial \alpha_s}+
\frac{2\alpha_s}{3\pi}\Big[2\ln\Big(\frac{\mu s}{s_0}\Big)+\mathcal{E}(x)\Big]\right)f^{(1)}_\Lambda(\mu)\eta_\Lambda(s,x,\mu)
\\
=
\frac{2\alpha_s}{3\pi}f^{(1)}_\Lambda(\mu) \int_{-\infty}^\infty dx'\, V(x,x')\,\eta_\Lambda(s,x',\mu)\,,
\end{multline}
where $\mathcal{E}(x)$ is defined in Eq.~(\ref{gamma}) and the kernel $V(x,x')$ is given by the matrix element
\begin{align}
\VEV{\phi_{s',x'}|\delta \mathcal{H}_{12}|\phi_{s,x}}=\delta(s-s') V(x,x')\,.
\end{align}
We obtain
\begin{equation}
V(x,x')=\frac1{2\pi}\sum_{n=0}^\infty \varkappa_n^{-1} \frac{H^{\ast}_n(x) H_n(x')}{(n+1)(n+2)}
=\frac1{2\sinh\pi(x-x')}\left[\frac{x-x'}{xx'}-\frac{\pi\sinh\pi(x-x')}{\sinh\pi x\sinh\pi x'}\right].
\end{equation}
It is easy to see that $V(x,x)\sim 1/(2\pi x^2)$ for large $x$, and decreases exponentially in
$|x-x'|$.

In order to solve (\ref{RGE-V}) one needs to find the eigenfunctions of the integral equation
\begin{align}
 \mathcal{E}(x)\,\eta_E(x)-[V \eta_E](x)= E\,\eta_E(x)\,.
\label{equation}
\end{align}
\begin{figure}[t]
\centerline{
\begin{picture}(280,170)(0,0)
\put(-5,0){
\includegraphics[width=10.0cm]{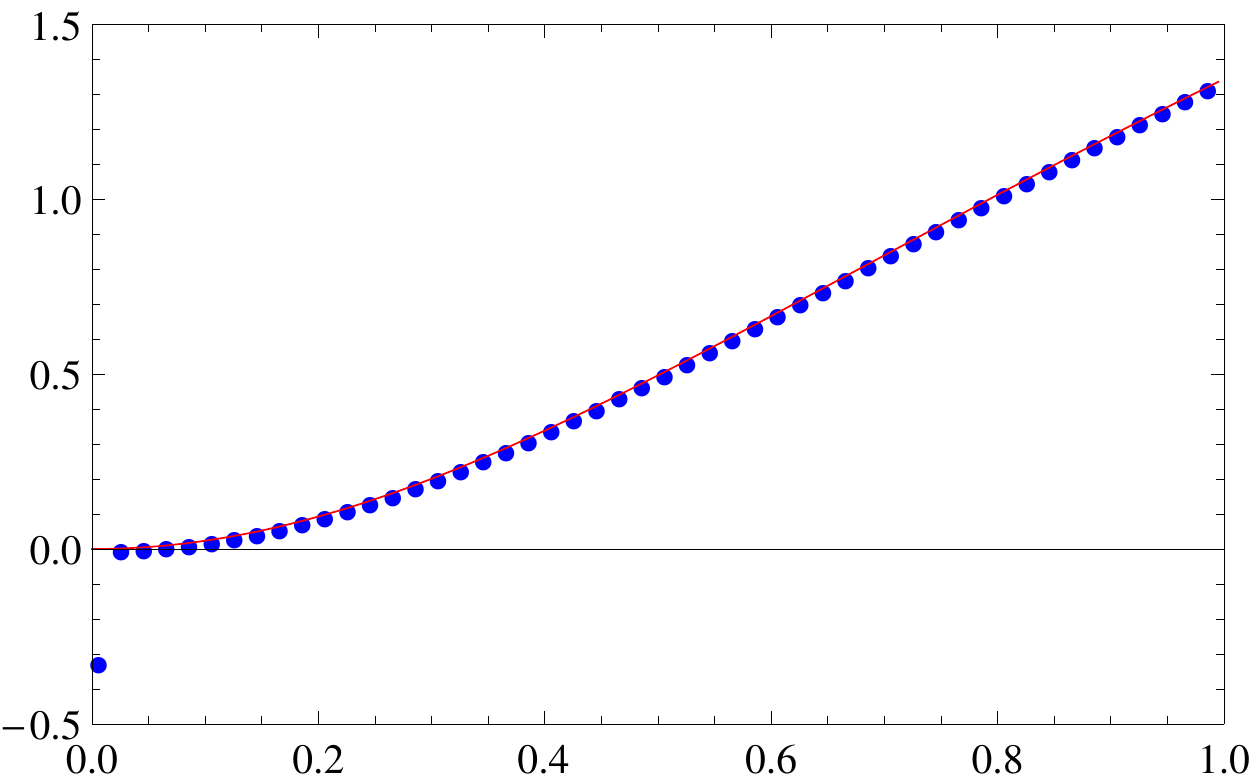}}
\put(135,-8){{\large $x_n$}}
\put(-20,70){{\large \rotatebox{90}{$E(x_n)$}}}
\end{picture}
}
\caption{\small
The spectrum of eigenvalues $E(x_n) \equiv E_n$ of the discretized version of Eq.~(\ref{equation})
$x\mapsto x_n = (2n+1)/200$, $n = 0,1,\ldots,99$ (blue dots) compared to
the ``unperturbed'' spectrum $\mathcal{E}(x_n)$ (red solid curve).
In order not to overload the plot, only every second eigenvalue is shown.
}
\label{fig1:spectrum}
\end{figure}
\begin{figure}[t]
\centerline{
\begin{picture}(280,220)(0,0)
\put(-5,0){
\includegraphics[width=10.0cm]{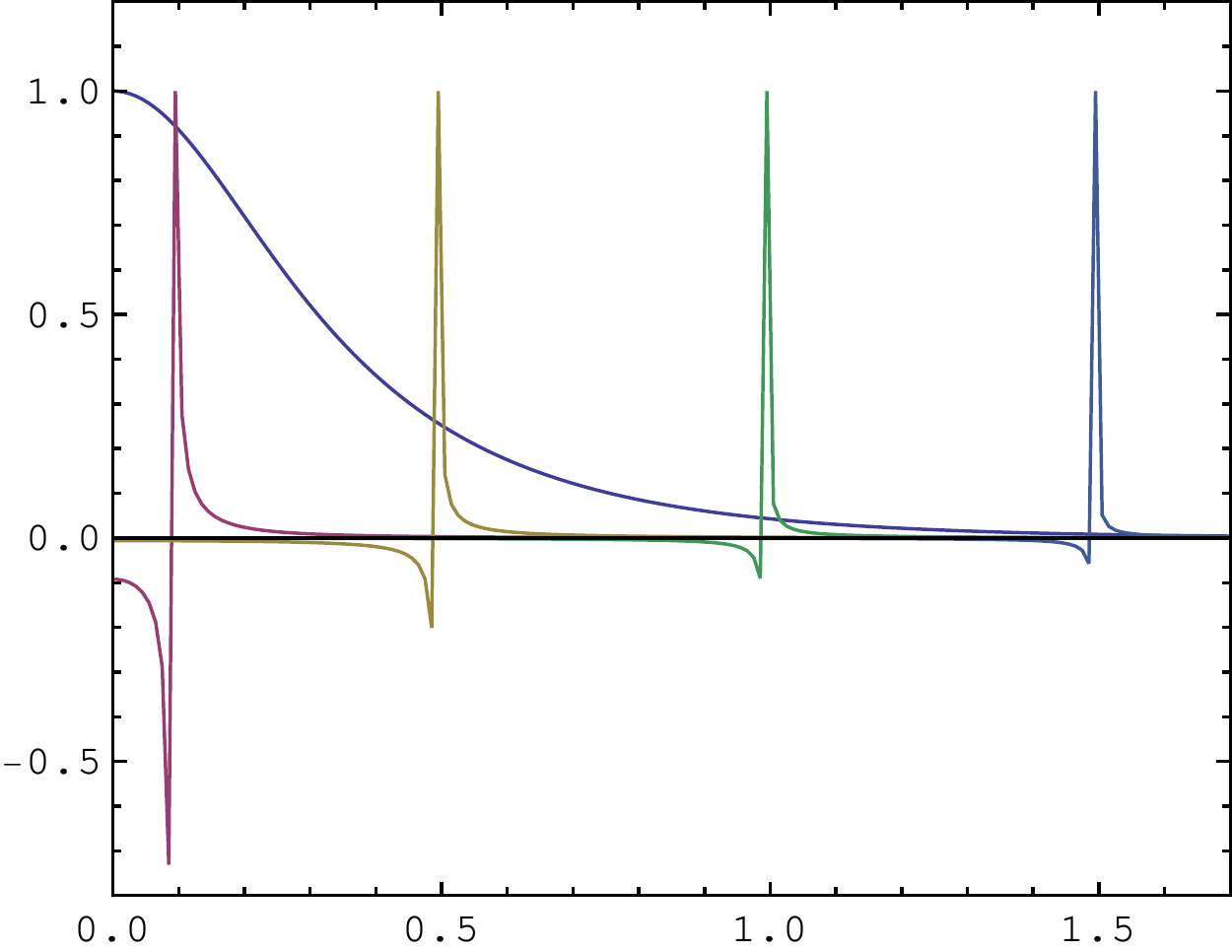}}
\put(135,-8){$x$}
\end{picture}
}
\caption{\small
The $x$-even eigenfunctions $\eta^+_k(x_n)=\eta^+_k(-x_n)$ for $k=0$ (the ground state), and $k=10, 50, 100, 150$
(from left to right), for $L=5$ and  $N=500$. Normalization is arbitrary.
}
\label{fig2:wf}
\end{figure}
If $V\to 0$, obviously all eigenfunctions are localized in $x$, $\eta_a(x) \sim \delta(x-a)$. The spectrum of
eigenvalues is continuous, $E_a = \mathcal{E}(a) \ge 0$, and double degenerate since
$\mathcal{E}(a) =  \mathcal{E}(-a)$.
In order to understand the effect of the ``perturbation'' $V$ we consider the discretized version of this
equation:  $x\to x_n=(n+1/2)\Delta x$, $\Delta x=L/N$, $n=-N,\ldots,N-1$.
The unperturbed eigenfunctions, $\eta_k(x_n)=\delta_{nk}$, correspond to local excitations at the $k-th$ site.
Discretizing the integral in (\ref{RGE-V}) one replaces the original eigenvalue problem (\ref{equation})
by the eigenvalue problem for the matrix $\mathcal{V}_{nk}=\delta_{nk} \mathcal{E}(x_n)-\Delta x\, V(x_n,x_m)$.
Since $V(x,x')=V(-x,-x')$, all eigenstates have definite parity with respect to $x\to -x$;
the double degeneracy is lifted and one can study $x$-even and $x$-odd eigenstates separately.
Diagonalising this matrix numerically we find that the shift of eigenvalues as compared to the unperturbed spectrum
is surprisingly small, $\delta E=E-\mathcal{E}\leq 0.003$, for all eigenstates except for the lowest one,
cf. Fig.~\ref{fig1:spectrum}, and the corresponding eigenfunctions $\eta_k(x_n)$ remain
well localized around the point $x_k$, see Fig.~\ref{fig2:wf}.
At the same time the lowest $x$-even eigenstate changes drastically: It becomes delocalized, see Fig.~\ref{fig2:wf}, and
separated from the rest of the spectrum by a finite gap%
\footnote{The size of the gap coincides with the gap in the spectrum of anomalous dimensions of
three-light-quark operators in the large-$N$ limit~\cite{Braun:1999te}, indicating that these problems
are related. Unravelling this connection goes beyond the tasks of this letter.}
\begin{align}
\Delta E=E_0\simeq-0.3214\,.
\label{gap}
\end{align}
In the continuum  limit $(\Delta x\to 0, L\to\infty)$ this phenomenon can be understood as creation of a bound
state in addition to the continuum spectrum that remains to be largely unperturbed.
The ``wave function'' of this (lowest) state can be approximated to a good accuracy (better that 1\% for $|x|<3$)
by the following expression:
\begin{align}\label{groundstate}
\eta_0(x)\simeq \frac{\sqrt{2}E_0}{\sqrt{2+x^2}}\frac{\rho(x)}{[E_0- \mathcal{E}(x)]}\,, && \eta_0(0)=1\,.
\end{align}
It can be convenient to expand this function in Hahn polynomials
\begin{align}
\eta_0(x) & =\sum_{n=0,2,\ldots}^{\infty} \chi_n \, H_{n}(x)\,,
\end{align}
where the first few coefficients read
\begin{align}
\chi_0\simeq0.612, &&\chi_2\simeq-0.126, && \chi_4\simeq0.0574, &&\chi_6\simeq-0.0338, && \chi_8\simeq0.0226, && \chi_{10}\simeq-0.0163\,.
\label{cn}
\end{align}
The normalization is given by
\begin{align}
\int_{-\infty}^\infty \frac{dx}{2\pi}\, \eta_0^2(x) = \sum_{n=0}^\infty \varkappa_n \, \chi_n^2 \simeq0.0758\,.
\end{align}
Coming back to the representation of the DA in the form (\ref{expandLambda}) we can separate the contribution
of the discrete (lowest) level as
\begin{align}
\eta_\Lambda(s,x,\mu)=\xi_0(s,\mu)\, \chi_0^{-1}\eta_0(x)+\eta^\prime_\Lambda(s,x,\mu).
\label{separate}
\end{align}
where the function $\eta^\prime_\Lambda(s,x,\mu)$ accounts for the contribution of the continuum spectrum
and must be orthogonal to $\eta_0(x)$,
\begin{align}
\int dx\,\eta_0(x)\,\eta^\prime_\Lambda(s,x,\mu)=0\,.
\label{ortho}
\end{align}
Going over to the momentum space  we obtain for the contribution of the lowest state (asymptotic DA)
\begin{eqnarray}
f^{(1)}_\Lambda(\mu) \widetilde \Psi^{(0)}_\Lambda(\omega,u;\mu)&=&
 f^{(1)}_\Lambda(\mu_0)  \chi_0^{-1}\omega^2 u\bar u \left(\frac{\mu}{\mu_0}\right)^{-\frac{8}{3\beta_0}}\!\! L^{\frac{4}{3\beta_0}[E_0-\frac{4\pi}{\beta_0\,\alpha_s(\mu_0)}]}
\!\int_{-\infty}^\infty\! \frac{dx}{2\pi} \eta_0(x)
\nonumber \\&&\times
 \int_0^\infty\!\!\!\! s ds\,
 \widetilde\phi_{s,x}(\omega,u)
\xi_0(s,\mu_0) \left(\frac{\mu_0 s}{s_0}\right)^{\frac{8}{3\beta_0}\ln L},
\label{asymptoticDA2}
\end{eqnarray}
cf. Eq.~(\ref{scaledep}). Note that the restriction to the contribution of the discrete level implies a
certain relation between the momentum fraction distribution between the two light quarks and their total momentum $\omega$,
the remaining freedom is encoded in the ``profile function'' $\xi_0(s,\mu_0)$ at the reference scale, which can be arbitrary.
For the simplest ansatz
\begin{align}
\xi_0(s,\mu_0)= s e^{-s \epsilon_0}\,,
\label{simple}
\end{align}
cf.~(\ref{etasx}), we obtain (at the scale $\mu_0$)
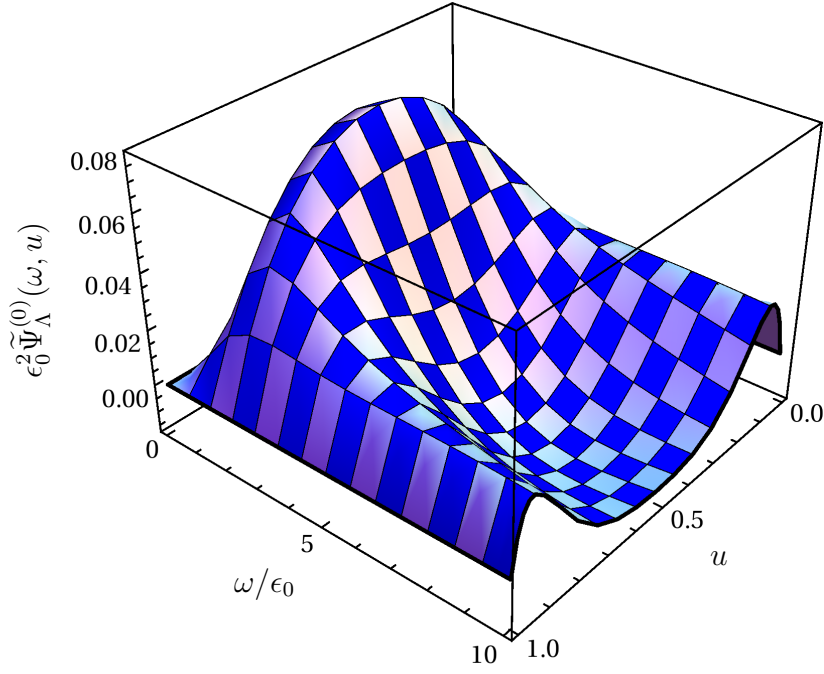
\begin{figure}[t]
\centerline{
\begin{picture}(280,220)(0,0)
\put(-5,-20){\includegraphics[width=10.0cm]{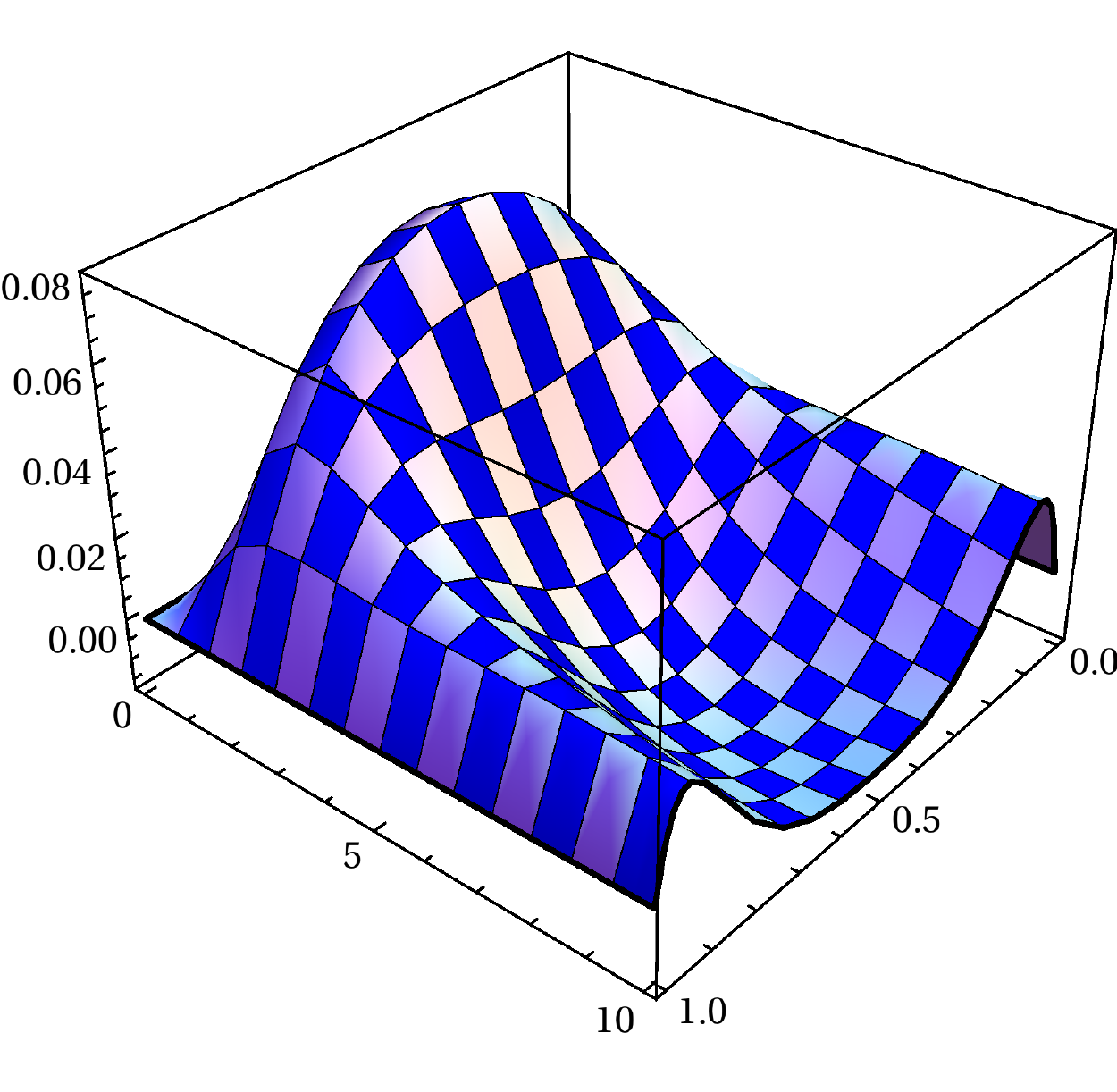}}
\put(57,22){{\large $\omega/\epsilon_0$}}
 \put(237,32){{\large $u$}}
\put(-28,103){{\large \rotatebox{90}{$\epsilon_0^2\widetilde \Psi^{(0)}_\Lambda(\omega,u)$}}}
\end{picture}
}
\caption{\small
 Asymptotic $\Lambda_b$ distribution amplitude for the simplest choice (\ref{simple}) of the profile
 function $\xi_0(s,\mu_0)$.
}
\label{fig3:3D}
\end{figure}
\begin{eqnarray}
 \widetilde\Psi^{(0)}_\Lambda(\omega,u) &=&
\frac1{\epsilon_0^2}  u\bar u\sum_{n=0,2,\ldots}^\infty  \,
i^n \left(\frac{\chi_n}{\chi_0}\right) \frac{\Gamma(n+4)}{\Gamma(2n+4)} C_n^{3/2}(1-2u)\, \left(\frac\omega{\epsilon_0}\right)^{n+2}
{}_1F_1\Big(\genfrac{}{}{0pt}{}{n+4}{2n+4}\Big|-\frac\omega{\epsilon_0}\Big)
\notag\\
   &=& 
  \frac1{\epsilon_0^2}  u\bar u \biggl\{  \left(\frac\omega{\epsilon_0}\right)^{2}
 {e^{-\omega/\epsilon_0}} 
- \frac{1}{42}
\left(\frac{\chi_2}{\chi_0}\right) \left(\frac\omega{\epsilon_0}\right)^{4} C_2^{3/2}(1-2u)\, 
{}_1F_1\Big(\genfrac{}{}{0pt}{}{6}{8}\Big|-\frac\omega{\epsilon_0}\Big)
+\ldots\biggr\},
\label{asymptoticDA}
\end{eqnarray}
where the first few Hahn expansion coefficients  $\chi_n$ are given in  Eq.~(\ref{cn}). The function
$\widetilde\Psi^{(0)}_\Lambda(\omega,u)$ is plotted in Fig.~\ref{fig3:3D}. Note that the contributions of higher Gegenbauer
polynomials $C_n^{3/2}(1-2u)$ in (\ref{asymptoticDA}) are accompanied by increasing powers of $\omega$ so that the
deviation from the "naive" $\sim u\bar u$ shape is increasing with the energy: the distribution becomes broader.
Taking into account that for large $\omega$
\begin{align}
 \omega^{n+2}
{}_1F_1\Big(\genfrac{}{}{0pt}{}{n+4}{2n+4}\Big|-\omega\Big)\simeq \omega^{-2}
\frac{\Gamma(2n+4)}{\Gamma(n)}\,, \qquad n> 0\,,
\label{1F1}
\end{align}
we see that all terms in this expansion except for the first one have a power-like asymptotics at $\omega\to \infty$.%
\footnote{
The exponential falloff of the first term is tightened to the particular choice of the profile function
(\ref{simple}) and in this sense accidental.}
One formally gets
\begin{align}\label{Aseta}
\widetilde\Psi^{(0)}_\Lambda(\omega,u)\underset{\omega\to\infty}{\simeq}
        \omega^{-2} u\bar u\sum_{n=2,4\ldots}^\infty i^n \left(\frac{\chi_n}{\chi_0}\right) \frac{\Gamma(n+4)}{\Gamma(n)}\,C_n^{3/2}(1-2u)\,.
\end{align}
The series in (\ref{Aseta}) is divergent indicating that for large $\omega$ the function
$\Psi^{(0)}_\Lambda(\omega,u)$ develops end-point singularities (at $u\to 0,1$). This feature is rather robust and does not
depend on the precise choice of the profile function $\xi_0(s,u)$ provided it vanishes sufficiently fast
at $s\to 0$ and $s\to \infty$.

We expect that the model of the $\Lambda_b$ DA in Eq.~(\ref{asymptoticDA}), or a more general one
in Eq.~(\ref{asymptoticDA2}), will be sufficient for phenomenological applications. If necessary, the contributions
of the continuum spectrum (\ref{separate}) can be added, in which case the scale dependence of $\eta^\prime_\Lambda(s,x,\mu)$
can be approximated using Eq.~(\ref{scaledep}). In this approximation
the orthogonality condition (\ref{ortho}) will not hold at all scales,
which is, however, unlikely to be numerically significant.

The evolution equation for the $\Lambda_b$ DA has also been discussed in Ref.~\cite{Bell:2013tfa} 
using a different representation $\eta_\Lambda(s,x) \mapsto \hat\rho_2(w_r,\varkappa)$ where we use the 
notatation $\varkappa$ for the variable called $u'$ in~\cite{Bell:2013tfa} to avoid confusion with the 
momentum fraction. The relation is simply $s = 1/w_r$ for the first variable, whereas going over from 
the $x$- to $\varkappa$-representation corresponds to the Fourier transform
\begin{align}
    \hat\eta(\varkappa) = \int_{-\infty}^{\infty} dx\, \eta(x) \left(\frac{\varkappa}{\bar\varkappa} \right)^{ix} =
     \int_{-\infty}^{\infty} dx\, \eta(x) e^{ipx} \,, \qquad p = \ln \frac{\varkappa}{\bar\varkappa}\,,\qquad \bar\varkappa = 1-\varkappa\,.    
\label{comparison}
\end{align}
In other words if our variable $x$ (\ref{CPsi}) is interpreted as a quasimomentum, then $p = \ln (\varkappa/\bar \varkappa)$ is the corresponding 
generalized coordinate%
\footnote{We thank the referee for suggesting to make this comparison.}.
For the ground state $\eta_0(x)=\eta_0(-x)$ implies $\hat\eta_0(\varkappa) = \hat\eta_0(1-\varkappa)$.

The end-point behavior of $\hat\eta_0(\varkappa)$ is determined by the 
position of the (nearest) singularity of $\eta_0(x)$ in the complex $x$ plane. A singularity (simple pole) at $x_0 = \pm i a$ corresponds to 
$\hat\eta_0(\varkappa) \sim [\varkappa\bar\varkappa]^a $ for $\varkappa\to 0, \varkappa\to 1$.
The position of the singularity can be related to the value of energy $E_0$, alias the lowest anomalous dimension.
One can show that the term in $V$ in Eq.~(\ref{equation}) does not contribute close to the singularity 
so that the following exact relation holds:
\begin{align}
  E_0 = \mathcal{E}(x_0) = \psi(1+a) + \psi(1-a) + 2\gamma_E\,.
\end{align}
Using an (approximate) value $E_0 = -0.3214$, Eq.~(\ref{gap}),  we obtain, to the same accuracy
\begin{align}
                    a = 0.3460\,. 
\label{u-sing}
\end{align}
Assuming that the asymptotic $\sim [\varkappa\bar\varkappa]^a $ behavior can be extrapolated to the whole interval $\varkappa \in [0,1]$
one obtains a model for the eigenfunction of the ground state
\begin{align}
      \hat\eta_0(\varkappa)   = [\varkappa\bar\varkappa]^a \qquad\mapsto\qquad \eta_0(x) = \Gamma[a+i x] \Gamma[a- ix] /\Gamma^2[a]\,, 
\label{alternative}
\end{align}
which turns out to be in good agreement numerically with a more complicated parametrization in Eq.~(\ref{groundstate}).

This result agrees well with the approximation for the asymptotic $\Lambda_b$ DA in the $\varkappa$-space 
$\hat\rho_2(w_r,\varkappa)\sim [\varkappa\bar\varkappa]^{1/3}$
found in Ref.~\cite{Bell:2013tfa} (see Fig.~4 there) by expanding the eigenfunction in Gegenbauer polynomials and retaining
the first few terms. Numerical convergence of this expansion (away from the end-points) observed in~\cite{Bell:2013tfa} 
is in fact directly related to our result that the lowest eigenstate of the evolution equation is separated from the continuum 
spectrum by a finite gap~(\ref{gap}).  

The above discussion cannot be directly applied to all other eigenstates of the evolution equation that belong
to the contunuum spectrum. In particular for the integrable case these solutions are given by plane waves, 
$\hat \eta(\varkappa)\sim e^{ipx}$, and possess an essential singularity at $\varkappa\to 0, \varkappa\to 1$. 
This singularity is only seen in extreme proximity to the end-points
and cannot be found using the Gegenbauer expansion. Whether such singularities are important for phenomenological applications remains to be studied.


\vspace*{0.5cm}
%

{\large \bf 5.}~~To summarize, in this letter we have studied the evolution equations that determine the scale
dependence of leading-twist DAs of heavy-light baryons containing one infinitely heavy and two light quarks.
The evolution equations are different for the cases that the light quarks have the same, or opposite, chirality.
For the first case, which corresponds to transverse DAs of $j^P=1^+$ sextets ($\Sigma_b,\Xi_b,\Omega_b$ and
$\Sigma^\ast_b,\Xi^\ast_b,\Omega^\ast_b$),
the evolution equation turns out to be completely integrable, that is it has a nontrivial integral of motion.
The anomalous dimensions form a continuum spectrum parametrized by two real numbers, $s$ and $x$ (\ref{gamma}),
and the corresponding eigenfunctions are known exactly (\ref{Psidef}),~(\ref{eigenfunctions}).
For the second case ($j^P=0^+$ $SU(3)_F$ triplet and all longitudinal DAs of heavy baryons in the
$j^P=1^+$ sextets), integrability is broken by an additional contribution to the evolution kernel that effectively
corresponds to an attractive interaction between the light quarks and creates a bound state.
As the result, the lowest anomalous  dimension becomes separated from the rest of the spectrum (that remains continuous)
by a finite gap (\ref{gap}). The corresponding eigenfunction is delocalized in the $x$-space and can be
found using numerical methods (\ref{groundstate}) (see also (\ref{alternative}). 
It can be interpreted as the asymptotic DA at large scales~(\ref{asymptoticDA2}),
(\ref{asymptoticDA}) and deviates
significantly from the naive $\sim u(1-u)$ shape at large quark energies, cf.~(\ref{Aseta}) and Fig.~\ref{fig3:3D}.

We expect that evolution equations for the higher-twist DAs of heavy baryons \cite{Ball:2008fw,Ali:2012pn} and for the three-particle
quark-gluon DAs of $B$-mesons in the large-$N_c$ limit~\cite{Knodlseder:2011gc} have similar properties and can be studied using
the same methods. These can be important for practical applications since heavy hadron decay form factors 
for physical values of the $b$-quark mass are likely to be dominated by soft contributions 
that can be related to higher-twist DAs using light-cone sum rules, see e.g.~\cite{Feldmann:2011xf,Braun:2012kp}.
  
Analogous unconventional integrable models with the Hamiltonian commuting with the diagonal entry $D(u)$ of the monodromy matrix
have appeared recently in the studies of high-energy scattering amplitudes in the $N=4$ supersymmetric Yang-Mills theory
\cite{Lipatov:2009nt,Bartels:2011nz,Basso:2010in,Belitsky:2011nn}.

%
\vspace*{0.5cm}
%

{\bf Acknowledgements}

This work was supported by RFBR grants 13-01-12405, 14-01-00341 (S.D.) and
by the DFG, grant BR2021/5-2 (A.M.).



\begin{thebibliography}{99}

\bibitem{Feldmann:2011xf}
  T.~Feldmann and M.~W.~Y.~Yip,
  Phys.\ Rev.\ D {\bf 85} (2012) 014035
   [Erratum-ibid.\ D {\bf 86} (2012) 079901].

\bibitem{Wang:2011uv}
  W.~Wang,
  Phys.\ Lett.\ B {\bf 708} (2012) 119.


\bibitem{Ball:2008fw}
  P.~Ball, V.~M.~Braun and E.~Gardi,
  Phys.\ Lett.\ B {\bf 665}, 197 (2008).

\bibitem{Ali:2012pn}
  A.~Ali, C.~Hambrock, A.~Y.~Parkhomenko and W.~Wang,
  Eur.\ Phys.\ J.\ C {\bf 73}, 2302 (2013).

\bibitem{Bell:2013tfa}
  G.~Bell, T.~Feldmann, Y.~-M.~Wang and M.~W.~Y.~Yip,
  JHEP {\bf 1311}, 191 (2013).

\bibitem{Braun:1998id}
  V.~M.~Braun, S.~E.~Derkachov and A.~N.~Manashov,
  Phys.\ Rev.\ Lett.\  {\bf 81}, 2020 (1998).

\bibitem{Braun:1999te}
  V.~M.~Braun, S.~E.~Derkachov, G.~P.~Korchemsky and A.~N.~Manashov,
  Nucl.\ Phys.\ B {\bf 553}, 355 (1999).


\bibitem{Lipatov:1993yb}
  L.~N.~Lipatov,
\emph{  ``High-energy asymptotics of multicolor QCD and exactly solvable lattice models,''}
  hep-th/9311037.

\bibitem{Lipatov:1994xy}
  L.~N.~Lipatov,
  JETP Lett.\  {\bf 59}, 596 (1994)
  [Pisma Zh.\ Eksp.\ Teor.\ Fiz.\  {\bf 59}, 571 (1994)].


\bibitem{Faddeev:1994zg}
  L.~D.~Faddeev and G.~P.~Korchemsky,
  Phys.\ Lett.\ B {\bf 342} (1995) 311.


\bibitem{Minahan:2002ve}
  J.~A.~Minahan and K.~Zarembo,
  JHEP {\bf 0303}, 013 (2003).

\bibitem{Beisert:2003tq}
  N.~Beisert, C.~Kristjansen and M.~Staudacher,
  Nucl.\ Phys.\ B {\bf 664}, 131 (2003).

\bibitem{Beisert:2003yb}
  N.~Beisert and M.~Staudacher,
  Nucl.\ Phys.\ B {\bf 670}, 439 (2003).

\bibitem{Faddeev:1996iy}
  L.~D.~Faddeev,
  {\it``How algebraic Bethe ansatz works for integrable model''},
  In: Quantum Symmetries/Symetries Quantiques,
  Proc. Les-Houches summer school, LXIV.
  Eds. A. Connes, K. Kawedzki, J. Zinn-Justin; North-Holland, 1998, 149-211,
  hep-th/9605187.

\bibitem{Kulish:1981bi}
  P.~P.~Kulish and E.~K.~Sklyanin,
  Lect.\ Notes Phys.\  {\bf 151} (1982) 61.

\bibitem{Sklyanin:1991ss}
  E.~K.~Sklyanin,
  {\it Quantum Inverse Scattering Method}, in
{\sl Quantum Groups and Quantum Integrable Systems}, (Nankai lectures), ed. Mo-Lin Ge, pp. 63-97, World Scientific
Publ., Singapore 1992, [hep-th/9211111]


\bibitem{Derkachov:2014gya}
  S.~E.~Derkachov and A.~N.~Manashov,
  \emph{Iterative construction of eigenfunctions of the monodromy matrix for SL(2,C) magnet,}
  arXiv:1401.7477 [math-ph].

\bibitem{Korchemsky:1991zp}
  G.~P.~Korchemsky and A.~V.~Radyushkin,
  Phys.\ Lett.\ B {\bf 279}, 359 (1992).


\bibitem{Lange:2003ff}
  B.~O.~Lange and M.~Neubert,
  Phys.\ Rev.\ Lett.\  {\bf 91}, 102001 (2003).

\bibitem{Braun:2003wx}
  V.~M.~Braun, D.~Y.~Ivanov and G.~P.~Korchemsky,
  Phys.\ Rev.\ D {\bf 69}, 034014 (2004).

\bibitem{Knodlseder:2011gc}
  M.~Knodlseder and N.~Offen,
  JHEP {\bf 1110}, 069 (2011).

\bibitem{Braun:2003rp}
  V.~M.~Braun, G.~P.~Korchemsky and D.~Mueller,
  Prog.\ Part.\ Nucl.\ Phys.\  {\bf 51}, 311 (2003).


\bibitem{Braun:2014owa}
  V.~M.~Braun and A.~N.~Manashov,
 Phys.\ Lett.\ B {\bf 731}, 316 (2014).

\bibitem{Bukhvostov:1985rn}
  A.~P.~Bukhvostov, G.~V.~Frolov, L.~N.~Lipatov and E.~A.~Kuraev,
  Nucl.\ Phys.\ B {\bf 258}, 601 (1985).

\bibitem{Derkachov:2002tf}
  S.~E.~Derkachov, G.~P.~Korchemsky and A.~N.~Manashov,
  JHEP {\bf 0307}, 047 (2003).


\bibitem{Braun:2012kp}
  V.~M.~Braun and A.~Khodjamirian,
  Phys.\ Lett.\ B {\bf 718} (2013) 1014.

\bibitem{Lipatov:2009nt}
  L.~N.~Lipatov,
  J.\ Phys.\ A {\bf 42} (2009) 304020.

\bibitem{Bartels:2011nz}
  J.~Bartels, L.~N.~Lipatov and A.~Prygarin,
  J.\ Phys.\ A {\bf 44} (2011) 454013.


\bibitem{Basso:2010in}
  B.~Basso,
  Nucl.\ Phys.\ B {\bf 857} (2012) 254.

\bibitem{Belitsky:2011nn}
  A.~V.~Belitsky,
  Phys.\ Lett.\ B {\bf 709} (2012) 280.




\end{thebibliography}
\end{document}